\documentclass[12pt,reqno]{article}
\usepackage{amsfonts}
\usepackage{amsmath}
\usepackage{amsbsy}

\usepackage{amssymb,latexsym}
\numberwithin{equation}{section}

\begin{document}
 \allowdisplaybreaks[1]
\title{On the Symmetric Space $\sigma$-model Kinematics}
\author{Nejat T. Y$\i$lmaz\\
Department of Mathematics
and Computer Science,\\
\c{C}ankaya University,\\
\"{O}\u{g}retmenler Cad. No:14,\quad  06530,\\
 Balgat, Ankara, Turkey.\\
          \texttt{ntyilmaz@cankaya.edu.tr}}
\maketitle
\begin{abstract}The solvable Lie
algebra parametrization of the symmetric spaces is discussed.
Based on the solvable Lie algebra gauge two equivalent
formulations of the symmetric space sigma model are studied. Their
correspondence is established by inspecting the normalization
conditions and deriving the field transformation laws.

\end{abstract}

\section{Introduction}
The symmetric space sigma model
\cite{west,tani,julia1,julia2,ker1,ker2,nej1,nej2} has a field
content of scalars which parametrize a homogeneous coset manifold
$G/K$ which is a Riemannian globally symmetric space for all the
$G$-invariant Riemannian structures on it \cite{hel}. The theory
is invariant under the parametrization preserving global (rigid)
$G$-action from the right and the local $K$-action from the left.
The scalars transform nonlinearly under these actions. In general
the global symmetry group $G$ is a non-compact real form of a
semi-simple Lie group and $K$ is its maximal compact subgroup. The
formulation of the symmetric space sigma model is based on the
Cartan-Maurer form which is induced by the local parametrization
of the coset representatives of $G/K$ by the scalar fields of the
theory. There are two equivalent formulations of the symmetric
space sigma model, the more conventional one which is also
applicable to the cases where the scalar manifold is not a
symmetric space is based on the decomposition of the Cartan-Maurer
form which reveals the vielbein and the gauge connection
identification of the scalar coset manifold $G/K$
\cite{west,tani,percaci}. This formulation divides the
Cartan-Maurer form into two parts; one is a coset generators
valued one-form and the other takes values in the algebra of the
maximal compact subgroup $K$. The second formulation of the
symmetric space sigma model is based on the introduction of an
internal metric \cite{julia1,julia2}. When one uses the solvable
Lie algebra parametrization \cite{nej2,fre} to generate the coset
representatives the analysis of the theory has simplifications in
both of the formulations \cite{ker1,ker2,nej1,nej2}. In
\cite{nej2} the Cartan-Maurer forms and the field equations of
both of the formulations are studied when the solvable Lie algebra
gauge is assumed.

In this work following the construction of the solvable Lie
algebra gauge or the solvable Lie algebra parametrization which
has an essential role in our analysis we will study the
lagrangians of both of the formulations in the solvable Lie
algebra gauge to find a correspondence between them. We will
inspect the normalization conditions then we will derive the
transformation laws between the field contents of both of the
formulations by assuming a normalization scheme. We will also
mention about the relation between the two formulations when the
model is coupled to other fields.

In section two we will introduce the solvable Lie algebra gauge to
parameterize the coset manifold $G/K$ of the symmetric space sigma
model. By using the solvable Lie algebra parametrization we will
derive the lagrangian of the symmetric space sigma model
explicitly for both of the above mentioned formulations in section
three. We will also compare the normalization conditions and we
will obtain the transformation laws between the field contents of
the two formulations by choosing a normalization convention.
\section{Symmetric Spaces and the Solvable Lie Algebra Parametrization}\label{section34}
The symmetric space sigma models are based on homogeneous
manifolds \cite{carter}, the homogeneity is in the sense that
there exists a transitive action of a Lie group $G$ on these
manifolds. These homogeneous spaces are in the form of a coset
manifold $G/K$ where $G$ is in general a non-compact real form of
any other semi-simple Lie group and $K$ is a maximal compact
subgroup of $G$. When $G$ is a compact real form its maximal
compact subgroup $K$ is $G$ itself thus the coset space $G/G$ is a
single point and the corresponding sigma model is an empty set.
The numerator group may as well be a split real form (maximally
non-compact) of a semi-simple Lie group.

The Lie algebra $k_{0}$ of the analytical subgroup $K$ is a
subalgebra of the semi-simple Lie algebra $g_{0}$ which is the Lie
algebra of $G$. The Lie algebra $k_{0}$ is a maximal compactly
imbedded Lie subalgebra of $g_{0}$ therefore it is an element of a
Cartan decomposition of $g_{0}$ \cite{hel}
\begin{equation}\label{ch341}
g_{0}=k_{0}\oplus p_{0},
\end{equation}
which is a vector space direct sum of the Lie subalgebra $k_{0}$
and a vector subspace $p_{0}$. Since $G$ is a linear analytical
Lie group the corresponding Lie algebra $g_{0}$ is non-compact for
both of the cases when $G$ is a non-compact or a split real form
(although we consider it separately we should not forget that it
is a limiting non-compact case). The map $u_{0}:k+p\longrightarrow
k-p$, for all $k\in k_{0}$ and $p\in p_{0}$ is the Cartan
involution which generates \eqref{ch341}. The pair $(g_{0},u_{0})$
is an orthogonal symmetric Lie algebra of the non-compact type
\cite{hel}. Thus we conclude that $(G,K)$ is associated with
$(g_{0},u_{0})$ and it is of non-compact type too. The Cartan
decomposition \eqref{ch341} is the eigenspace decomposition of
$u_{0}$ where the elements of $k_{0}$ have $+1$ eigenvalues and
the elements of $p_{0}$ have $-1$ eigenvalues under the involution
$u_{0}$. We also know that $(G,K)$ is a Riemannian symmetric pair
therefore the coset space $G/K$ has a unique analytical structure
induced by the quotient topology of $G$. The scalar manifold $G/K$
is a Riemannian globally symmetric space for all the $G$-invariant
Riemannian structures on $G/K$.\footnote{This is the origin of the
name; symmetric space sigma model.} The crucial consequence of
this identification is that the exponential map
$Exp:g_{0}\longrightarrow G$ induces a diffeomorphism
\begin{equation}\label{ch342}
Exp:p_{0}\longrightarrow G/K,
\end{equation}
from the ${\Bbb{R}}^{dimp_{0}}$ manifold $p_{0}$ onto $G/K$ since
it maps the elements of $p_{0}$ onto the representatives of the
left cosets $G/K$ \cite{hel}. This result will enable us to define
a parametrization of the scalar manifold $G/K$ on which the sigma
model will be constructed in the next section.

Furthermore one may use the Iwasawa decomposition of $g_{0}$ which
is built on the Cartan decomposition \eqref{ch341} \footnote{In
this respect the Iwasawa decomposition is not a Cartan
decomposition but it introduces a solvable Lie algebra of $g_{0}$
which is isomorphic to $p_{0}$ thus which generates another
parametrization of $G/K$ via \eqref{ch342} \cite{hel}.} and one
may make use of the root space decomposition basis of $g_{0}$ to
parametrize the scalar coset manifold. The Iwasawa decomposition
reads
\begin{equation}\label{ch343}
\begin{aligned}
g_{0}&=k_{0}\oplus s_{0}\\
\\
&=k_{0}\oplus h_{p_{0}}\oplus n_{0},
\end{aligned}
\end{equation}
where $k_{0}$ is the Lie algebra of $K$ and the algebra direct sum
$s_{0}=h_{p_{0}}\oplus n_{0}$ is a solvable Lie subalgebra of
$g_{0}$ which is isomorphic to the vector space $p_{0}$. In
\eqref{ch343} $h_{p_{0}}$ is generated by $r$ non-compact Cartan
generators $\{H_{i}\}$. Also the nilpotent Lie subalgebra $n_{0}$
in \eqref{ch343} is generated by a subset $\{E_{\beta}\}$ of the
positive root generators of $g_{0}$ where
$\beta\in\Delta_{nc}^{+}$. The roots in $\Delta_{nc}^{+}$ are the
non-compact roots with respect to a Cartan involution $\theta$
which is composed of the conjugation induced by the Cartan
decomposition in \eqref{ch343} and the conjugation of $g_{0}$ via
its complexification \cite{hel}. The Cartan subalgebra $h_{0}$
generates an abelian subgroup in $G$ which is called the torus.
Although we call it torus it is not the ordinary torus
topologically in fact it has the topology $(S^{1})^{m}\times
\mathbb{R}^{n}$ for some $m$ and $n$ and if it is diagonalizable
in $\mathbb{R}$ (such that $m=0$) then it is called an $R$-split
torus. These definitions can be generalized for the subalgebras of
$h_{0}$ as well. The subspace of $G$ which is generated by
$h_{p_{0}}$ is the maximal $R$-split torus in $G$ in the sense
defined above and its dimension is called the $R$-rank which we
will denote by $r$. If $r$ is maximal such that $r=l$ where $l$ is
the rank of $G$ ($l=$dim$h_{0}$), which also means that
$h_{p_{0}}=h_{0}$ then the Lie group $G$ is said to be in split
real form (maximally non-compact). In this case $h_{p_{0}}=h_{0}$
is generated by all the Cartan generators $\{H_{i}\}$ and
$\Delta_{nc}^{+}=\Delta^{+}$ so that the generators
$\{E_{\beta}\}$ of $n_{0}$ correspond to the entire set of
positive roots. Thus the solvable Lie subalgebra $s_{0}$ coincides
with the Borel subalgebra which is generated by the entire Cartan
and the positive root generators of $g_{0}$ for the split real
form case. If on the other hand $r$ is minimal such that $r=0$
then $G$ is a compact real form. All the other cases in between
are called non-compact semi-simple real forms.

For the non-compact real form $G$ if we consider the Iwasawa
decomposition and in \eqref{ch342} if we use the basis
$\{H_{i},E_{\beta}\}$ which generates the solvable Lie subalgebra
$s_{0}$ then we have the parametrization
\begin{equation}\label{ch344}
Exp:\sum{\Bbb{R}}\{H_{i},E_{\beta}\}\longrightarrow G/K.
\end{equation}
\eqref{ch344} is called the solvable Lie algebra parametrization
or the solvable Lie algebra gauge of the symmetric space $G/K$
\cite{fre}. On the other hand when $g$ is a split real form
(maximally non-compact) if we use the Borel subalgebra basis which
is made up of the entire set of the Cartan generators and the
positive root generators; then \eqref{ch344} is called the Borel
parametrization or the Borel gauge of $G/K$.

In summary in this section we have obtained a legitimate
parametrization of the symmetric space $G/K$ by using the solvable
Lie subalgebra $s_{0}$ of $g_{0}$. If we use the notation
$\{T_{m}\}$ for the basis vectors $\{H_{j},E_{\beta}\mid
j=1,...,r\; ;\;\beta\in\Delta_{nc}^{+}\}$ of $s_{0}$ and if
$\{\varphi^{m}(x)\}$ are $C^{\infty}$-maps over the
$D$-dimensional spacetime then the map
\begin{equation}\label{ch345}
  \nu(x)=e^{\varphi^{m}(x)T_{m}},
\end{equation}
is an onto $C^{\infty}$-map from the $D$-dimensional spacetime to
the Riemannian  globally symmetric space $G/K$. The gauge map,
\eqref{ch345} which depends on the scalar functions
$\{\varphi^{m}(x)\}$ is the building block in the construction of
the symmetric space sigma model.

\section{Normalization Conditions and the Duality Transformations of the SSSM}
In this section we will obtain the field transformations of the
two equivalent formulations of the symmetric space sigma model
(SSSM) which are based on the solvable Lie algebra parametrization
introduced in the last section. We will show that when one assumes
a normalization convention relating the matrix representations of
the basis that is used in \eqref{ch345} of the two separate
formulations one can find a correspondence between the sets of
field definitions of the two distinct constructions.

In order to construct the symmetric space sigma model we first
consider the set of $G$-valued maps $\nu(x)$. They transform onto
each other as $\nu\rightarrow k(x)\nu g$, $\forall g\in G$,
$k(x)\in K$ for some subgroup $K$ of $G$. We will assume that the
map $\nu(x)$ corresponds to a parametrization of the coset $G/K$
(for convenience we will consider the left cosets). Thus as
mentioned before we assume that the map $\nu(x)$ is from the
$D$-dimensional spacetime into the group $G$ and its range is
composed of the representatives of the left cosets of $G/K$.
Moreover if $G$ is a non-compact real form of a semi-simple Lie
group and $K$ is a maximal compact subgroup of $G$ then $G/K$
becomes a symmetric space and $\nu(x)$ can be taken as the map
\eqref{ch345} such that $\nu(x)=e^{\varphi^{m}(x)T_{m}}$ by using
the Cartan and the Iwasawa decompositions as we have mentioned in
the last section. In this case the transformation rule
$\nu\rightarrow k(x)\nu g$, $\forall g\in G$, $k(x)\in K$ which we
assign on $\nu(x)$ preserves the gauge based on the Iwasawa
decomposition.

A lagrangian which is invariant under the transformations
discussed above can be given as
\begin{equation}\label{ch4117}
 {\mathcal{L}}_{1}=\frac{c}{4}\, tr(\ast d{\mathcal{M}}^{-1}\wedge
 d{\mathcal{M}}),
\end{equation}
where the internal metric ${\mathcal{M}}$ is defined as
${\mathcal{M}}=\nu ^{\#}\nu $ and $c$ is a constant which may
arise when the symmetric space sigma model is coupled to other
fields. Here $\#$ is the generalized transpose over the Lie group
$G$ such that $(exp(g))^{\#}=exp(g^{\#})$. It is induced by the
Cartan involution $\theta$ over $g_{0}$ ($g^{\#}=-\theta(g)$
$\forall g\in g_{0}$) \cite{hel}. As mentioned in \cite{ker1} it
is possible to find a matrix representation of the Lie algebra
$g_{0}$ in which $\#$ coincides with the matrix transpose
operator. For this reason one can define an induced $\#$ map over
the group $G$ as $(exp(g))^{\#}=exp(g^{\#})$. If the subgroup of
$G$ generated by the compact generators is an orthogonal group
then in the fundamental representation of $g_{0}$ the generators
can be chosen such that $g^{\#}=g^{T}$ for $g\in g_{0}$. Also if
the subgroup of $G$ generated by the compact generators is a
unitary group then in the fundamental representation
$g^{\#}=g^{\dagger}$ for $g\in g_{0}$. In our formulation we will
assume that we choose a representation in which $g^{\#}=g^{T}$.
The lagrangian \eqref{ch4117} can be expressed in terms of the
pullback of the Cartan-Maurer form
\begin{equation}\label{ss1}
{\mathcal{G}}=d\nu \nu ^{-1},
\end{equation}
as follows
\begin{subequations}\label{ch4118}
\begin{align}
{\mathcal{L}}_{1}&=\frac{c}{4}\, tr(\ast d{\mathcal{M}}^{-1}\wedge
 d{\mathcal{M}})\notag\\
\notag\\
&=\frac{c}{4}\, tr(-\ast d\nu \nu ^{-1}\wedge (d\nu \nu ^{-1})^{\#}+\ast d\nu^{-1}\nu \wedge \nu ^{-1}d\nu\notag\\
\notag\\
&\quad+ \ast d(\nu^{\#})^{-1}\nu^{\#} \wedge
(\nu^{\#})^{-1}d\nu^{\#}+ (\ast \nu d\nu^{-1} \wedge
(\nu^{-1})^{\#}d\nu^{\#})^{\#})\notag\\
\notag\\
&=\frac{c}{4}\, tr(-\ast d\nu \nu ^{-1}\wedge (d\nu \nu ^{-1})^{\#}-\ast d\nu\nu^{-1} \wedge d\nu\nu^{-1}\notag\\
\notag\\
&\quad+ \ast\nu d\nu^{-1} \wedge d\nu \nu^{-1}- \ast d\nu \nu^{-1}
\wedge (d\nu\nu^{-1})^{\#})\notag\\
\notag\\
&=-\frac{c}{2}\, tr(\ast \mathcal{G}\wedge \mathcal{G}^{\#}+\ast
\mathcal{G}\wedge \mathcal{G}) .\tag{\ref{ch4118}}
\end{align}
\end{subequations}
In the above derivation we have made use of the identities
\begin{equation}\label{ss2}
\nu ^{-1}d\nu=-d\nu ^{-1}\nu\quad ,\quad d\nu\nu ^{-1}=-\nu d\nu
^{-1}.
\end{equation}
Also by bearing in mind the definition of $\nu$ we have used the
properties of the transpose operation such as
\begin{equation}\label{ss3}
(\nu^{-1})^{\#}=(\nu^{\#})^{-1}\quad,\quad
tr(\nu^{\#})=tr(\nu)\quad ,\quad
 (\nu_{1}\nu_{2})^{\#}=\nu_{2}^{\#}\nu_{1}^{\#}\quad ,\quad (\nu^{\#})^{\#}=\nu.
\end{equation}
Cyclic permutations are always permissible under the trace
operator however one must be careful about the fact that if
$\nu_{1}$ and $\nu_{2}$ are matrix valued functions under the
representation chosen then
\begin{equation}\label{ss4}
tr(d\nu_{1}\wedge \ast d\nu_{2})=(-1)^{(D-1)}tr(\ast
d\nu_{2}\wedge d\nu_{1}),
\end{equation}
where $D$ is the dimension of the spacetime. In the derivation of
\eqref{ch4118} we have also used that $(d\nu)^{\#}=d(\nu)^{\#}$.
Now we will explicitly calculate $\mathcal{G}$. By using the
matrix identity
\begin{equation}\label{ss5}
de^{C}e^{-C}=dC+\frac{1}{2!}[C,dC]+\frac{1}{3!}[C,[C, dC]]+....,
\end{equation}
we obtain
\begin{equation}\label{ss6}
\begin{aligned}
\mathcal{G}&=de^{\varphi^{i}T_{i} }e^{-\varphi^{i}T_{i}}\\
\\
&=d\varphi^{i}T_{i} +\frac{1}{2!}[\varphi^{i}T_{i}
,d\varphi^{j}T_{j} ] +\frac{1}{3!}[\varphi^{i}T_{i}
,[\varphi^{j}T_{j} ,d\varphi^{k}T_{k} ]]+....\\
\\
&=d\varphi^{i}T_{i}
+\frac{1}{2!}\varphi^{i}d\varphi^{j}C_{ij}^{k}T_{k}
+\frac{1}{3!}\varphi^{i}\varphi^{j}d\varphi^{k}C_{jk}^{l}C_{il}^{t}T_{t}+....\\
\\
&=\overset{\rightharpoonup }{
\mathbf{T}}\:\mathbf{\Delta}\:\overset{\rightharpoonup }{\mathbf{
d\varphi}}.
\end{aligned}
\end{equation}
We have defined
\begin{equation}\label{ss7}
[T_{m},T_{n}]=C_{mn}^{k}T_{k}.
\end{equation}
Also the components of the row vector $\overset{\rightharpoonup }{
\mathbf{T}}$ are $T_{i}=H_{i}$ for $i=1,...,r$ and
$T_{\alpha+r}=E_{\alpha}$ for $\alpha=1,...,$dim$n_{0}$.
$\overset{\rightharpoonup }{\mathbf{d\varphi}}$ is a column vector
of the field strengths $\{d\varphi^{m}\}$ and the
dim$s_{0}\times$dim$s_{0}$ matrix $\mathbf{\Delta}$ can be given
as
\begin{equation}\label{ss8}
\begin{aligned}
\mathbf{\Delta}&=\sum\limits_{n=0}^{\infty }\dfrac{M
^{n}}{(n+1)!}\\
\\
&=(e^{M}-I)M^{-1},
\end{aligned}
\end{equation}
 where $M_{m}^{n}=\varphi^{l}C_{lm}^{n}$. If we
 insert the explicit form of $\mathcal{G}$ calculated in
 \eqref{ss6} in the lagrangian \eqref{ch4118} we obtain
\begin{equation}\label{ss9}
\mathcal{L}_{1}=-\frac{c}{2}\Delta^{n}_{l}\ast d\varphi^{l}\wedge
\Delta^{m}_{k}d\varphi^{k}(tr(T_{n}T^{\#}_{m})+tr(T_{n}T_{m})) .
\end{equation}
If we assume that the scalar maps when coupled to the generators
in \eqref{ch345} generate ranges in a sufficiently small
neighborhood around the identity element of $g_{0}$ then we can
introduce another parametrization of the coset elements as a
product of two exponential maps
\begin{equation}\label{ss10}
\nu (x)=e^{\frac{1}{2}\phi ^{i}(x)H_{i}}e^{\chi
^{\beta}(x)E_{\beta}},
\end{equation}
where the fields $\{\phi^{i}\}$ are called the dilatons and
$\{\chi^{\beta}\}$ are called the axions. The maps in \eqref{ss10}
and \eqref{ch345} can be chosen to be equal to each other if the
locality condition of the range is assumed as discussed above
\cite{hel,carter,wybourne,satinger,onis}. In \cite{nej1,nej2} the
Cartan form $\mathcal{G}$ is calculated in terms of the fields
$\{\phi^{i}, \chi^{\beta}\}$ as
\begin{equation}\label{ss11}
\mathcal{G}=\frac{1}{2}d\phi ^{i}H_{i}+\overset{\rightharpoonup }{
\mathbf{E}^{\prime }}\:\mathbf{\Omega }\:\overset{\rightharpoonup
}{\mathbf{d\chi}},
\end{equation}
where we have defined the row vector $(\overset{\rightharpoonup }{
\mathbf{E}^{\prime }})_{\alpha}=e^{%
\frac{1}{2}\alpha _{i}\phi ^{i}}E_{\alpha}$ and
$\overset{\rightharpoonup }{\mathbf{d\chi}}$ is a column vector of
the field strengths of the axions $\{d\chi^{\beta}\}$. Also
$\mathbf{\Omega}$ is a dim$n_{0}\times$dim$n_{0}$ matrix
\begin{equation}\label{ss12}
\begin{aligned}
 \mathbf{\Omega}&=\sum\limits_{n=0}^{\infty }\dfrac{\omega
^{n}}{(n+1)!}\\
\\
&=(e^{\omega}-I)\,\omega^{-1}.
\end{aligned}
\end{equation}
The matrix $\omega$ is defined as $\omega _{\beta }^{\gamma }=\chi
^{\alpha }\,K_{\alpha \beta }^{\gamma }$ where we define the
structure constants $K_{\alpha \beta }^{\gamma }$ and the root
vector components $\alpha_{i}$ as
\begin{equation}\label{ss13}
[E_{\alpha },E_{\beta }]=K_{\alpha \beta }^{\gamma }\,E_{\gamma
}\quad ,\quad [H_{i},E_{\alpha}]=\alpha_{i}\,E_{\alpha }.
\end{equation}
By using \eqref{ss11} in \eqref{ch4118} we can express the
lagrangian ${\mathcal{L}}_{1}$ in terms of the fields $\{\phi^{i},
\chi^{\beta}\}$ as
\begin{subequations}\label{ss14}
\begin{align}
{\mathcal{L}}_{1}&=-\frac{c}{8}\,\ast d{\phi}^{i}\wedge
 d\phi^{j}(tr(H_{i}H_{j}^{\#})+tr(H_{i}H_{j}))\notag\\
\notag\\
&\quad -\frac{c}{4}\,\ast d{\phi}^{i}\wedge
e^{\frac{1}{2}\alpha_{i}\phi^{i}} F^{\alpha}(tr(H_{i}E_{\alpha}^{\#})+tr(E_{\alpha}H_{i}^{\#})+
tr(H_{i}E_{\alpha})+tr(E_{\alpha}H_{i}))\notag\\
\notag\\
&\quad -\frac{c}{2}\,e^{\frac{1}{2}\alpha_{i}\phi^{i}}\ast
F^{\alpha}\wedge e^{\frac{1}{2}\beta_{i}\phi^{i}}
F^{\beta}(tr(E_{\alpha}E_{\beta}^{\#})+tr(E_{\alpha}E_{\beta})),\tag{\ref{ss14}}
\end{align}
\end{subequations}
where we have defined
\begin{equation}\label{ss15}
F^{\alpha}=\mathbf{\Omega}^{\alpha}_{\beta}d\chi^{\beta}.
\end{equation}
If we compare the lagrangian we have obtained in \eqref{ss14} with
the one
\begin{equation}\label{ss16}
\mathcal{L=}-\frac{1}{2}\sum\limits_{i=1}^{r}\ast d\phi ^{i}\wedge d%
\phi ^{i}-\frac{1}{2}\sum\limits_{\alpha\in
\Delta_{nc}^{+}}e^{\alpha _{i}\phi ^{i}}\ast F^{\alpha }\wedge
F^{\alpha },
\end{equation}
which is given in \cite{ker1,nej1} and whose equations of motion
are also derived therein we find that the normalization conditions
for the matrix representatives of the basis elements $\{H_{i},
E_{\alpha}\}$ must be in the form
\begin{subequations}
\begin{gather}\label{ss17}
tr(H_{i}H_{j}^{\#})+tr(H_{i}H_{j})=\frac{4}{c}\delta_{ij},\notag\\
\notag\\
tr(H_{i}E_{\alpha}^{\#})+tr(E_{\alpha}H_{i}^{\#})+
tr(H_{i}E_{\alpha})+tr(E_{\alpha}H_{i})=0,\notag\\
\notag\\
tr(E_{\alpha}E_{\beta}^{\#})+tr(E_{\alpha}E_{\beta})=\frac{1}{c}\delta_{\alpha\beta}\tag{\ref{ss17}}.
\end{gather}
\end{subequations}
As already given in \cite{nej2} if we compare \eqref{ss11} with
\eqref{ss6} we can find the transformation relations between the
field strengths $\{d\varphi^{m}\}$ and
$\{d\phi^{i},d\chi^{\beta}\}$ as
\begin{equation}\label{ss18}
\begin{aligned}
\mathbf{\Delta}_{m}^{i}d\varphi^{m}&=\frac{1}{2}d\phi^{i},\\
\\
\mathbf{\Delta}_{m}^{\beta+r}d\varphi^{m}&=e^{\frac{1}{2}\beta_{j}\phi^{j}}\mathbf{\Omega}_{\rho}^{\beta}
d\chi^{\rho},
\end{aligned}
\end{equation}
where the indices above are $i,j=1,...,r$;
$\beta,\rho=1,...,$dim$n_{0}$; $ m=1,...,$dim$s_{0} =r+$dim$n_{0}$
and we should remark that we enumerate the roots
$\beta\in\Delta_{nc}^{+}$.

As we have discussed in the introduction there exits another
formulation of the symmetric space sigma model. By using another
set of fields $\{\varphi^{\prime m}\}$ let us consider the coset
element
\begin{equation}\label{ss19}
  \nu^{\prime}(x)=e^{\varphi^{\prime m}(x)T_{m}}.
\end{equation}
We can define the Cartan form
\begin{equation}\label{ss20}
{\mathcal{G}^{\prime}}=\nu^{\prime -1}d\nu ^{\prime}=P+Q,
\end{equation}
where
\begin{equation}\label{ss21}
P=P^{m}T_{m}\quad,\quad Q=Q^{n}K_{n},
\end{equation}
$\{K_{n}\}$ being the generators of $k_{0}$ which is an element of
the Cartan decomposition of $g_{0}$ in \eqref{ch341}. In general
in the gauge \eqref{ss19} the coset generators do not have to form
up a subalgebra. Thus the Cartan form \eqref{ss20} can have
components both in $P$ and $Q$ directions since the algebra
product of the coset generators may result in the set $\{K_{n}\}$.
However when we take the coset generators $\{T_{m}\}$ to be the
generators of the solvable Lie algebra $s_{0}$ of the Iwasawa
decomposition \eqref{ch343} we have a simplification. As we will
explicitly calculate in the following lines in this case we have
\begin{equation}\label{ss22}
Q=Q^{n}K_{n}=0\quad \text{and}\quad\mathcal{G}^{\prime}=P.
\end{equation}
This is owing to the fact that the coset generators being the
generators of the solvable Lie algebra in this special gauge are
not mapped on the generators $\{K_{n}\}$ under the algebra
product. If we assume a representation in which the generators
$\{K_{n}\}$ and $\{T_{m}\}$ of the Iwasawa decomposition
\eqref{ch343} are orthogonal namely
\begin{equation}\label{ss22.5}
tr(k_{0}s_{0})=0,
\end{equation}
then the Iwasawa decomposition \eqref{ch343} satisfies the
requirements of the decomposition needed to build up the
equivalent vielbein formulation of the symmetric space sigma model
which is invariant under the global $G$ and the local $K$
transformations thus an invariant lagrangian can be given as
\cite{west,tani}\footnote{Even if a representation which enables
\eqref{ss22.5} does not exist the transformation between the
fields which we will derive below justifies the definition of the
lagrangian \eqref{ss23}. Therefore the introduction of
\eqref{ss23} can legitimately be considered as a redefinition of
fields.}
\begin{equation}\label{ss23}
{\mathcal{L}}_{2}=c^{\prime}\, tr(\ast P\wedge P).
\end{equation}
We may explicitly calculate this lagrangian by calculating the
Cartan form \eqref{ss20}. By using the matrix identity
\begin{equation}\label{ss24}
e^{-C}de^{C}=dC-\frac{1}{2!}[C,dC]+\frac{1}{3!}[C,[C, dC]]-....,
\end{equation}
one can calculate $\mathcal{G}^{\prime}$ as
\begin{equation}\label{ss25}
\begin{aligned}
\mathcal{G}^{\prime}&=e^{-\varphi^{\prime i}T_{i} }de^{\varphi^{\prime i}T_{i}}\\
\\
&=d\varphi^{\prime i}T_{i} -\frac{1}{2!}[\varphi^{\prime i}T_{i}
,d\varphi^{\prime j}T_{j} ] +\frac{1}{3!}[\varphi^{\prime i}T_{i}
,[\varphi^{\prime j}T_{j} ,d\varphi^{\prime k}T_{k} ]]-....\\
\\
&=d\varphi^{\prime i}T_{i} -\frac{1}{2!}\varphi^{\prime
i}d\varphi^{\prime j}C_{ij}^{k}T_{k}
+\frac{1}{3!}\varphi^{\prime i}\varphi^{\prime j}d\varphi^{\prime k}C_{jk}^{l}C_{il}^{t}T_{t}-....\\
\\
&=\overset{\rightharpoonup }{
\mathbf{T}}\:\mathbf{W}\:\overset{\rightharpoonup }{\mathbf{
d\varphi^{\prime}}}.
\end{aligned}
\end{equation}
We have defined the dim$s_{0}\times$dim$s_{0}$ matrix $\mathbf{W}$
as
\begin{equation}\label{ss26}
\begin{aligned}
\mathbf{W}&=\sum\limits_{n=0}^{\infty }\dfrac{(-1)^{n}\, M
^{\prime n}}{(n+1)!}\\
\\
&=(I-e^{-M^{\prime}})M^{\prime -1},
\end{aligned}
\end{equation}
 where $M_{m}^{\prime n}=\varphi^{\prime l}C_{lm}^{n}$.
We can now write the lagrangian \eqref{ss23} as
\begin{equation}\label{ss26.5}
\mathcal{L}_{2}=c^{\prime}W^{n}_{l}\ast d\varphi^{\prime l}\wedge
W^{m}_{k}d\varphi^{\prime k}tr(T_{n}T_{m}).
\end{equation}
As discussed before one can also express the coset map
$\nu^{\prime}$ in terms of the dilatons $\phi^{\prime i}$ and the
axions $\chi^{\prime\alpha}$ as
\begin{equation}\label{ss27}
\nu^{\prime} (x)=e^{\frac{1}{2}\phi ^{\prime i}(x)H_{i}}e^{\chi
^{\prime \beta}(x)E_{\beta}}.
\end{equation}
In \cite{nej2} in terms of the fields $\{\phi^{\prime i},
\chi^{\prime\alpha}\}$ the Cartan form $\mathcal{G}^{\prime}$ is
calculated as
\begin{equation}\label{ss28}
\mathcal{G}^{\prime}=\frac{1}{2}d\phi ^{\prime
i}H_{i}+\overset{\rightharpoonup}{\mathbf{E }}\:\mathbf{\Sigma
}\:(\overset{\rightharpoonup }{U}+\overset{\rightharpoonup
}{d\chi^{\prime}}),
\end{equation}
where the components of the row vector $(\overset{\rightharpoonup
}{ \mathbf{E}})_{\alpha}$ are $E_{\alpha}$ and
\begin{equation}\label{ss29}
U^{\alpha}=\frac{1}{2}\chi ^{\prime\alpha}\alpha_{i} d\phi
^{\prime i}.
\end{equation}
Also the dim$n_{0}\times$dim$n_{0}$ matrix $\mathbf{\Sigma}$ is
\begin{equation}\label{ss30}
 \mathbf{\Sigma}=\sum\limits_{n=0}^{\infty }\dfrac{(-1)^{n}\omega
^{\prime n}}{(n+1)!},
\end{equation}
with $\omega _{\beta }^{\prime\gamma }=\chi ^{\prime\alpha
}\,K_{\alpha \beta }^{\gamma }$. Now if we compare \eqref{ss25}
with \eqref{ss28} we find the transformation laws between the
fields $\{\varphi^{\prime m}\}$ and  $\{\phi^{\prime i},
\chi^{\prime\beta}\}$ as
\begin{equation}\label{ss31}
\begin{aligned}
\mathbf{W}_{m}^{i}d\varphi^{\prime m}&=\frac{1}{2}d\phi^{\prime i},\\
\\
\mathbf{W}_{m}^{\alpha +r}d\varphi^{\prime
m}&=\mathbf{\Sigma}_{\beta}^{\alpha} (U^{\beta}+d\chi^{\prime
\beta}),
\end{aligned}
\end{equation}
where $i=1,...,r$; $\alpha,\beta=1,...,$dim$n_{0}$; and $
m=1,...,$dim$s_{0} =r+$dim$n_{0}$.

Now under a special set of normalization conditions we may find
the field transformation laws between the primed and the unprimed
fields. To find a straightforward transformation rule after
inspecting the two lagrangians in \eqref{ss9} and \eqref{ss26.5}
we conclude that the normalization conventions in the first
(unprimed) formulation and the second (primed) formulation should
be chosen such that
\begin{equation}\label{ss36}
 (tr(T_{n}T_{m}))_{2}=-\frac{c}{2c^{\prime}}(tr(T_{n}T_{m})+tr(T_{n}T_{m}^{\#}))_{1},
\end{equation}
where the subscripts in $(tr(\:\:\:))_{1}$ and $(tr(\:\:\:))_{2}$
correspond to the separate trace conventions for the matrix
representatives we choose for the two formulations. Under the
above normalization conventions the comparison of \eqref{ss9} with
\eqref{ss26.5} now denotes that
\begin{equation}\label{ss37}
\mathcal{L}_{1}=\mathcal{L}_{2},
\end{equation}
if we identify
\begin{equation}\label{ss38}
\varphi^{m}=-\varphi^{\prime m}.
\end{equation}
This is obvious since if \eqref{ss38} is chosen then
\begin{equation}\label{ss39}
\mathcal{G}=-\mathcal{G^{\prime}}\quad\text{and}\quad
\mathbf{W}=\mathbf{\Delta}.
\end{equation}
Therefore from \eqref{ss18} we have
\begin{equation}\label{ss40}
\begin{aligned}
-\mathbf{\Delta}_{m}^{i}d\varphi^{\prime m}&=\frac{1}{2}d\phi^{i},\\
\\
-\mathbf{\Delta}_{m}^{\beta+r}d\varphi^{\prime
m}&=e^{\frac{1}{2}\beta_{j}\phi^{j}}\mathbf{\Omega}_{\alpha}^{\beta}
d\chi^{\alpha}.
\end{aligned}
\end{equation}
Finally comparing this result with \eqref{ss31} and using
\eqref{ss39} leads us to the transformation relations between
$\{\phi^{i},\chi^{\alpha}\}$ and $\{\phi^{\prime i},\chi^{\prime
\alpha}\}$ which we wish to find
\begin{equation}\label{ss41}
\begin{aligned}
d\phi^{i}&=-d\phi^{\prime i},\\
\\
e^{\frac{1}{2}\beta_{j}\phi^{j}}\mathbf{\Omega}_{\alpha}^{\beta}
d\chi^{\alpha}&=-\mathbf{\Sigma}_{\rho}^{\beta}
(U^{\rho}+d\chi^{\prime \rho}).
\end{aligned}
\end{equation}
Before concluding as a final remark we will mention about the
relevance of the two formulations when the symmetric space sigma
model is coupled to other fields $F^{i}=dA^{i}$. The interaction
lagrangian in this case can be given as
\cite{julia1,julia2,ker2,nej3,heterotic}
\begin{equation}\label{ss42}
\begin{aligned}
 \mathcal{L}_{m}&=-\frac{c_{m}}{2}\mathcal{M^{\prime}}_{kl} F^{k}\wedge\ast
 F^{l}\\
\\
&=-\frac{c_{m}}{2}F\wedge\mathcal{M^{\prime}} \ast F,
\end{aligned}
\end{equation}
where we define
\begin{equation}\label{ss43}
\begin{aligned}
 \mathcal{M^{\prime}}= \nu^{\prime\#}\nu^{\prime}.
\end{aligned}
\end{equation}
The total lagrangian in the second formulation of the symmetric
space sigma model then becomes
\begin{equation}\label{ss44}
{\mathcal{L}}_{coup}^{\prime}=c^{\prime}\, tr(\ast P\wedge
P)-\frac{c_{m}}{2}F\wedge\mathcal{M^{\prime}} \ast F.
\end{equation}
As we have mentioned before if the subgroup of $G$ generated by
the compact generators is an orthogonal group then in the
fundamental representation of $g_{0}$ the generators can be chosen
such that $g^{\#}=g^{T}$ for $g\in g_{0}$. The orthogonal global
symmetry groups obey this property thus if $G=O(m,n)$ then in the
fundamental representation we can take
$(\:\:\:\:)^{\#}=(\:\:\:\:)^{T}$. One additional condition can be
introduced for the coset elements $\nu^{\prime}$ and $\nu$. As it
can be explicitly seen in \cite{d=7,d=8} and also effectively used
in \cite{nej4,nej5,nej6} when $G=O(m,n)$ one can choose a basis in
which the coset representatives are locally represented by
symmetric matrices so that we can assume locally
\begin{equation}\label{ss45}
\nu^{\prime T}=\nu^{\prime}\quad,\quad \nu^{T}=\nu.
\end{equation}
When we choose our representation as above under the
transformation law \eqref{ss38} we have
\begin{equation}\label{ss46}
\mathcal{M}=\eta\mathcal{M^{\prime}}\eta,
\end{equation}
where $\eta$ is the indefinite signature metric which takes part
in the definition of the orthogonal group $G=O(m,n)$. In general
$\eta$ is a symmetric matrix which has $m$ positive and $n$
negative eigenvalues. If $A\in O(m,n)$ then
\begin{equation}\label{ss46.5}
A^{T}\eta A=\eta.
\end{equation}
In writing \eqref{ss46} we have also used the fact that
$\eta^{-1}=\eta$. Now if one assumes the transformation law
\eqref{ss38} one can use the lagrangian
\begin{equation}\label{ss47}
{\mathcal{L}}_{coup}=\frac{c}{4}\, tr(\ast
d{\mathcal{M}}^{-1}\wedge
 d{\mathcal{M}})-\frac{c_{m}}{2}F^{\prime}\wedge\mathcal{M}
\ast F^{\prime},
\end{equation}
instead of \eqref{ss44} which becomes equal to the former under
the prescribed duality transformations. In \eqref{ss47} we have
also performed the transformation
\begin{equation}\label{ss48}
F^{\prime}=\eta F,
\end{equation}
on the gauge field strengths  assuming the usual dimensional
matching between the number of the gauge fields and the dimension
of the representation chosen for the coset representatives which
is the dimension of the fundamental representation of $O(m,n)$.
When one performs the above-mentioned replacement for the
lagrangians one should inspect the coexistence of the
normalization conventions thus the transformation laws we have
derived earlier with the local symmetry conditions \eqref{ss45} of
the representation chosen.
\section{Conclusion}
Since our analysis is based on the solvable Lie algebra gauge we
have started with a discussion about how one can construct this
gauge for the symmetric spaces. Then by assuming the solvable Lie
algebra parametrization of the scalar coset, for two equivalent
formulations of the symmetric space sigma model we have derived
the lagrangians explicitly. We have compared the normalization
conditions and obtained the duality transformation laws for the
field contents of these separate formulations. Thus in this work
we have not only studied the two equivalent formulations of the
symmetric space sigma model in detail but we have also obtained
the correspondence of their field contents by deriving the
transformation laws between them. In addition we have extended the
duality transformations to include the matter fields when the
symmetric space sigma model has matter couplings.

The majority of the scalar sectors of the supergravity theories
with or without matter multiplets are formulated by symmetric
space sigma models \cite{sssugradivdim}. However the scalar
lagrangians are generally expressed in the second formulation
mentioned in section three. Therefore one needs to find out the
transformation laws between the field contents to express the
supersymmetry transformation laws if the first formulation is
used. By prescribing a relation between the normalization
conditions of the two constructions we have obtained the duality
transformation laws. Since the supersymmetry requires certain
coefficients for the scalar lagrangians when they are coupled to
other fields we have taken general coefficients in our derivation.
Our formulation also establishes the transformation rules between
two separate parametrizations of the coset elements in either of
the constructions.

Although we have assumed a normalization convention for deriving
the field transformations we have not questioned the explicit
representations which may obey these normalization conditions.
This issue can separately be studied in general terms or for
specific examples of the supergravity theories. In constructing
the representations which obey the normalization convention that
we have introduced one has the degrees of freedom which the
solvable Lie algebra parametrization provides. This is due to the
fact that our analysis is performed in a general and
representation free formalism from the algebraic point of view. In
this respect one has the freedom of choosing the Cartan
decomposition, the root space decomposition, the Cartan
subalgebra, the solvable Lie algebra, the basis, also the
representation used for the Lie algebra of the global symmetry
group $G$. Another comparison can also be separately studied which
relates the solvable Lie algebra gauge field content and the most
general coset parametrization whose Cartan-Maurer form has also a
piece in the $K$ generators.

\end{document}